\documentclass[amsmath,amssymb,lengthcheck,onecolumn,aps,prl] {revtex4}
\usepackage[T1]{fontenc}
\usepackage[utf8]{inputenc}
\usepackage{graphics}
\usepackage{graphicx}

\usepackage{xcolor}

\begin{document}
\title{\textbf{Dual-Space Analysis of Vortex Configurations and Rotational Energy in a 2D Optical Lattice}}

\author{Avra Banerjee\footnote{avra.pcs@bose.res.in}}

\affiliation{Department of Physics of Complex Systems, 
S. N. Bose National Centre for Basic Sciences, 
Block JD, Sector III, Salt Lake, Kolkata 700106, India}

\begin{abstract}
We investigate the effects of a stationary two-dimensional periodic optical lattice on a rotating Bose--Einstein condensate confined by an isotropic harmonic trap. Within the mean-field description, the condensate is modeled by the two-dimensional Gross--Pitaevskii equation in the rotating frame and solved numerically using an imaginary-time split-step Crank--Nicolson method to obtain stationary vortex states. We examine how the lattice depth, lattice wave vector, particle number, and angular velocity influence the vortex configurations, real-space density, momentum-space density, and energetics of the condensate. We find that the stationary periodic lattice significantly modifies vortex nucleation and spatial organization: vortices emerge progressively with increasing angular velocity and, for sufficiently strong lattice strength, undergo rearrangement and merging. Importantly, changes in the vortex configuration are directly reflected in the momentum-space density, demonstrating that vortex nucleation and rearrangement produce pronounced modifications of the condensate momentum distribution. We further analyze the energy per particle and rotational energy as functions of the angular velocity and find that the presence of the optical lattice substantially alters the rotational response of the condensate. In particular, the rotational energy is progressively reduced at higher angular velocities as the lattice strength is increased. These results demonstrate how a stationary periodic optical lattice modifies vortex formation, momentum-space structure, and rotational energetics of a rotating trapped Bose--Einstein condensate.

\end{abstract}

\maketitle

\section*{Introduction}

Bose--Einstein condensation in dilute ultracold atomic gases provides a well-controlled platform for investigating quantum many-body phenomena at ultralow temperatures. The macroscopic properties of a dilute condensate can be described within the mean-field approximation by a complex order parameter $\psi(\mathbf{r},t)$, whose dynamics are governed by the Gross--Pitaevskii (GP) equation. The interaction strength between the atoms can be controlled experimentally through the $s$-wave scattering length, for example, by means of a Feshbach resonance \cite{Anthony J. Leggett,fesh,fesh1,fesh2}.

A rotating Bose--Einstein condensate supports quantized vortices, which arise from the phase winding of the condensate order parameter \cite{rot_bec1,rot_bec2,rot_bec3,rot_bec4,rot_bec5}. Vortices in atomic condensates have been experimentally generated and observed using different techniques, including stirring, rapid rotation, and phase imprinting \cite{set1,set2,set3,rapid1,rapid2,phase1,phase2,phase3}. The presence of vortices strongly influences the density, phase structure, angular momentum, and energetics of the condensate. Rotating condensates can exhibit a variety of vortex-related phenomena, including vortex-lattice formation, vortex merging, collective vortex excitations, and other strongly correlated vortex states \cite{vort_lat,vort_amal,osci,bending,qhall,melting, BA_vort,luo_vortex_2026,dong_vortex_2026}. Consequently, the response of a rotating condensate to external spatial potentials is an important problem in the study of inhomogeneous quantum fluids.

Periodic optical potentials generated by interfering laser beams provide an experimentally controllable way of modifying the spatial structure of an ultracold atomic gas \cite{opp1,opp2}. Such a periodic potential is conventionally referred to as an optical lattice. Its depth and spatial periodicity can be controlled through the intensity and wavelength of the laser fields, respectively. Optical lattices have been used to investigate a wide range of phenomena in ultracold gases, including Bloch oscillations, tunneling, Josephson dynamics, superfluidity, and number squeezing \cite{op1,op2,op3,op4,op5,op6,op7}. When a rotating condensate is subjected to an optical lattice, the lattice introduces an additional spatial energy scale that competes with the harmonic confinement, interatomic interaction, and rotational energy. This competition can modify the formation and spatial arrangement of vortices.

The behavior of vortices in rotating condensates in periodic optical lattices has been investigated previously, including studies of vortex-lattice structures, vortex rearrangement, and lattice-induced effects on rotating condensates \cite{ROL1,ROL2,TROL3,ROL4,ROL5,ROL6,ROL7,ROL8}. In particular, the effects of periodic potentials on vortex configurations and vortex pinning have been studied in detail in earlier theoretical works \cite{reff1,reff2}. These studies establish that the presence of a periodic potential can substantially modify the vortex structure and energetics of a rotating condensate. Therefore, the effect of an optical lattice on a rotating BEC is not, by itself, a new problem. The motivation of the present work is instead to examine, within a common numerical framework, how the lattice parameters and rotation frequency influence the vortex configurations and the rotational energetics of the condensate for the parameter range considered here.

\begin{figure*}
\includegraphics[width=0.6\textwidth]{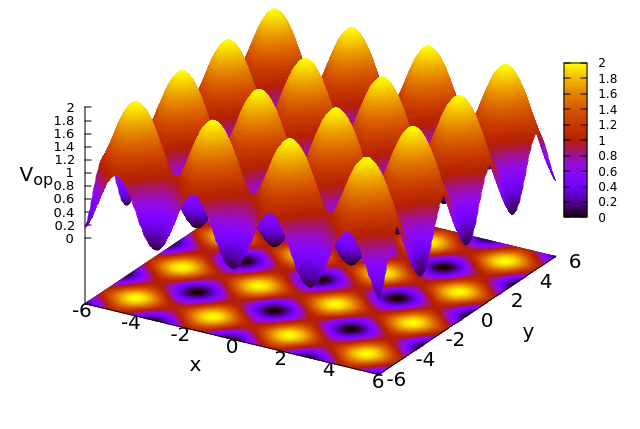}
\caption{Two-dimensional square optical lattice potential $V_{\rm op}(x,y)$ for lattice depth $V_0=1$ and wave vector $k=1$. The potential is repulsive and periodic along both the $x$ and $y$ directions. The spatial coordinates $x$ and $y$ are expressed in units of the characteristic length $l_0$.}
\label{op}
\end{figure*}

In the present study, we consider a two-dimensional rotating Bose--Einstein condensate confined by a harmonic trap and subjected to a stationary repulsive square optical lattice. The optical lattice is fixed in the laboratory frame and is described by the periodic potential
\begin{equation}
V_{\rm op}(x,y)=V_0\left[\sin^2(kx)+\sin^2(ky)\right],
\end{equation}
where $V_0$ determines the lattice depth and $k$ determines its spatial periodicity. The condensate is described by the two-dimensional Gross--Pitaevskii equation in the rotating frame and is solved numerically using an imaginary-time split-step Crank--Nicolson method to obtain stationary condensate configurations.

Our primary aim is to investigate how the stationary periodic lattice modifies the vortex configurations, momentum-space density, and energetic response of the rotating condensate. We examine the dependence of the condensate structure on the number of particles $N$, lattice depth $V_0$, lattice wave vector $k$, and angular velocity $\omega$. In particular, we compare the real-space vortex configurations and corresponding momentum-space density distributions in the absence and presence of the optical lattice. We focus on the changes in vortex nucleation \cite{nucle}, spatial rearrangement, and merging induced by the lattice, and examine how these changes are reflected in the momentum-space distribution. We further analyze the energy per particle and rotational energy as functions of the angular velocity.

For the parameter regime investigated, we find that the stationary optical lattice substantially modifies the sequence and spatial organization of vortex formation compared with the lattice-free condensate. At intermediate lattice strengths, vortex formation occurs progressively with increasing angular velocity, whereas stronger lattice potentials lead to pronounced vortex rearrangement and merging. The corresponding momentum-space density is also strongly modified by the changes in the vortex configuration, providing a clear Fourier-space signature of the lattice-induced restructuring of the condensate. We further find that the rotational energy is reduced relative to the lattice-free case, with the reduction becoming progressively more pronounced as the optical-lattice depth $V_0$ is increased, particularly at higher angular velocities. These results provide a numerical characterization of the interplay between rotation and a stationary periodic potential in a harmonically confined Bose--Einstein condensate, and demonstrate how the lattice strength can modify both the real-space vortex structure and its momentum-space manifestation.

\section*{Model and calculation}

We consider a two-dimensional Bose--Einstein condensate (BEC) confined in an isotropic harmonic trap and subjected to a square periodic optical lattice. The optical lattice can be generated by two mutually perpendicular standing-wave laser fields. For the configuration considered here, the corresponding lattice potential is given by \cite{vortex1}

\begin{equation}
    V_{\rm op}(x,y)
    =
    V_0\left[\sin^2(kx)+\sin^2(ky)\right],
    \label{Vlat}
\end{equation}

where $V_0$ denotes the lattice depth and $k$ is the magnitude of the wave vector of the standing waves. The corresponding lattice spacing is

\begin{equation}
    d=\frac{\pi}{k}.
    \label{lattice_spacing}
\end{equation}

Thus, $V_0$ determines the strength of the periodic modulation, whereas $k$ controls its spatial periodicity. The two standing waves can be arranged such that their mutual interference is suppressed, for example by using orthogonal polarizations or an appropriate relative phase \cite{vortex1}. The lattice parameters can therefore be controlled experimentally.

The periodic structure of the lattice introduces an additional spatial energy scale into the trapped condensate. In the present work, we are interested in how this periodic modulation modifies the density distribution, vortex configurations, momentum-space distribution, and rotational energetics of a rotating condensate.

\subsection*{Gross--Pitaevskii equation}

Within the mean-field approximation, the dynamics of the rotating condensate are described by the Gross--Pitaevskii (GP) equation. In the dimensionless units used throughout this work, the GP equation in the frame rotating about the $z$ axis with angular frequency $\omega$ is written as \cite{GPE1,GPE8}
\begin{figure*}
    \includegraphics[width=1.05\textwidth]{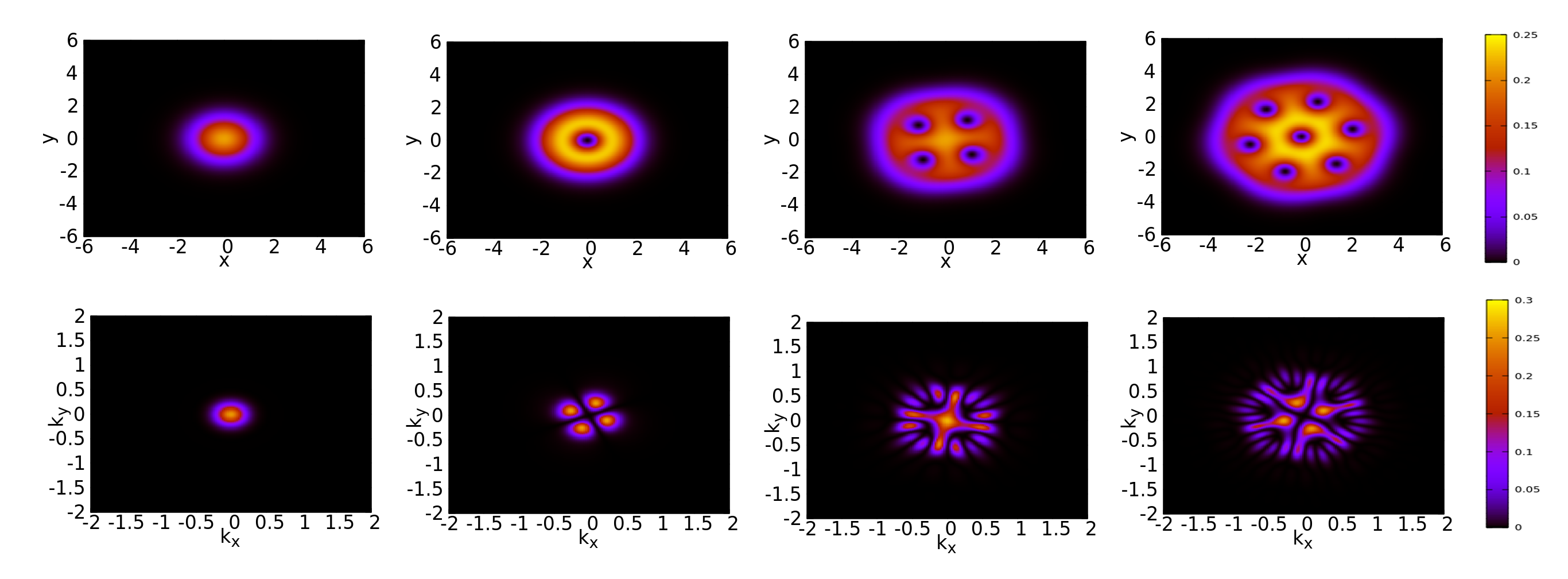}
    \caption{
    Real-space density (upper row) and corresponding momentum-space
    density (bottom row) of the rotating condensate in the absence of
    the optical lattice for different particle numbers:
    $N=500$ (first column), $N=2000$ (second column),
    $N=5000$ (third column), and $N=10000$ (fourth column).
    Here, $\omega=0.8$. The number and spatial arrangement of vortices
    change with the particle number, leading to corresponding changes
    in the momentum-space density. The momentum-space distribution is
    particularly sensitive to changes in the vortex configuration.
    The coordinates $x$ and $y$ are expressed in units of $l_0$, while
    $k_x$ and $k_y$ are expressed in units of $l_0^{-1}$. The units of
    $|\psi(x,y)|^2$ and $|\phi(k_x,k_y)|^2$ are $l_0^{-2}$ and $l_0^2$,
    respectively.
    }
    \label{vary_N}
\end{figure*}

\begin{figure*}
    \includegraphics[width=1.03\textwidth]{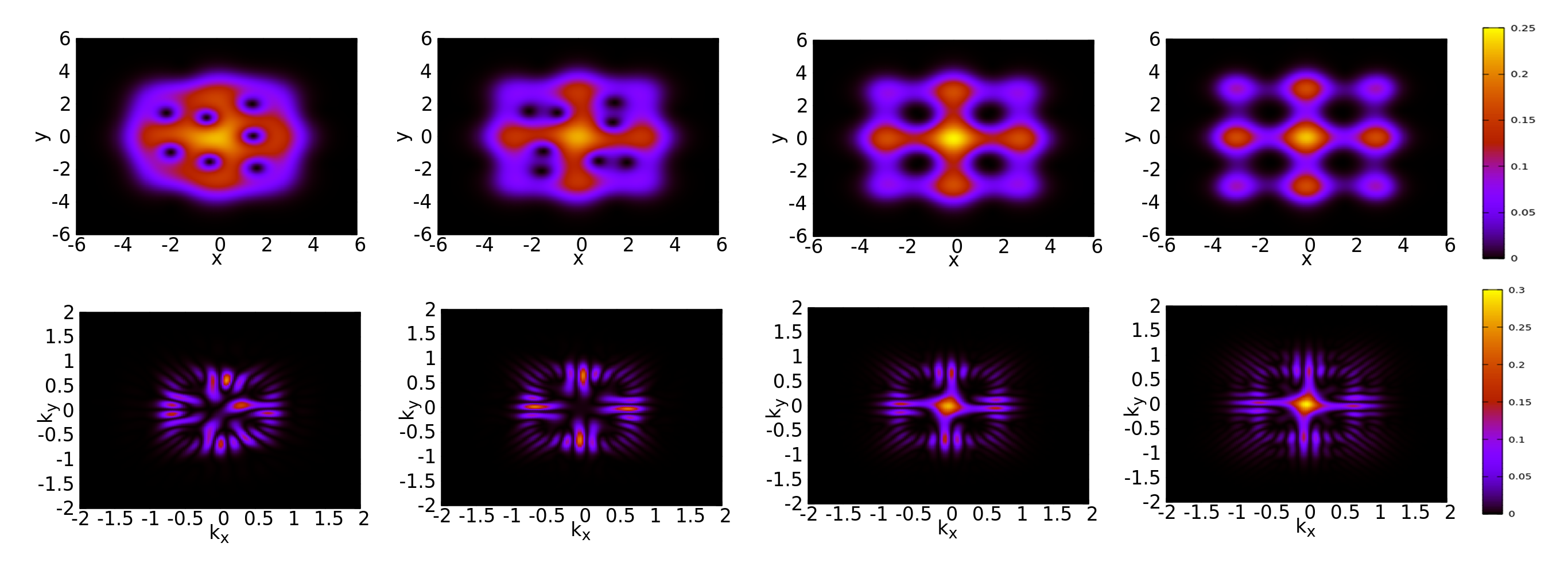}
    \caption{
    Real-space density (upper row) and corresponding momentum-space
    density (bottom row) for different optical-lattice strengths:
    $V_0=1$ (first column), $V_0=3$ (second column),
    $V_0=5$ (third column), and $V_0=10$ (fourth column).
    Here, $N=10000$, $\omega=0.8$, and $k=1$.
    The vortices become increasingly ordered as the lattice strength
    increases. Seven vortices are observed for $V_0=1$, whereas eight
    vortices are obtained for $V_0=3$ with a more ordered spatial
    arrangement. For larger lattice strengths, vortices merge to form
    four larger vortex cores, and the vortex structure is progressively
    suppressed for sufficiently large $V_0$. The corresponding
    momentum-space density also changes with the lattice strength,
    reflecting the modifications of the vortex configuration in real
    space. The coordinates $x$ and $y$ are expressed in units of $l_0$,
    while $k_x$ and $k_y$ are expressed in units of $l_0^{-1}$. The units
    of $|\psi(x,y)|^2$ and $|\phi(k_x,k_y)|^2$ are $l_0^{-2}$ and $l_0^2$,
    respectively.
    }
    \label{vary_v}
\end{figure*}

\begin{figure*}
    \includegraphics[width=1.03\textwidth]{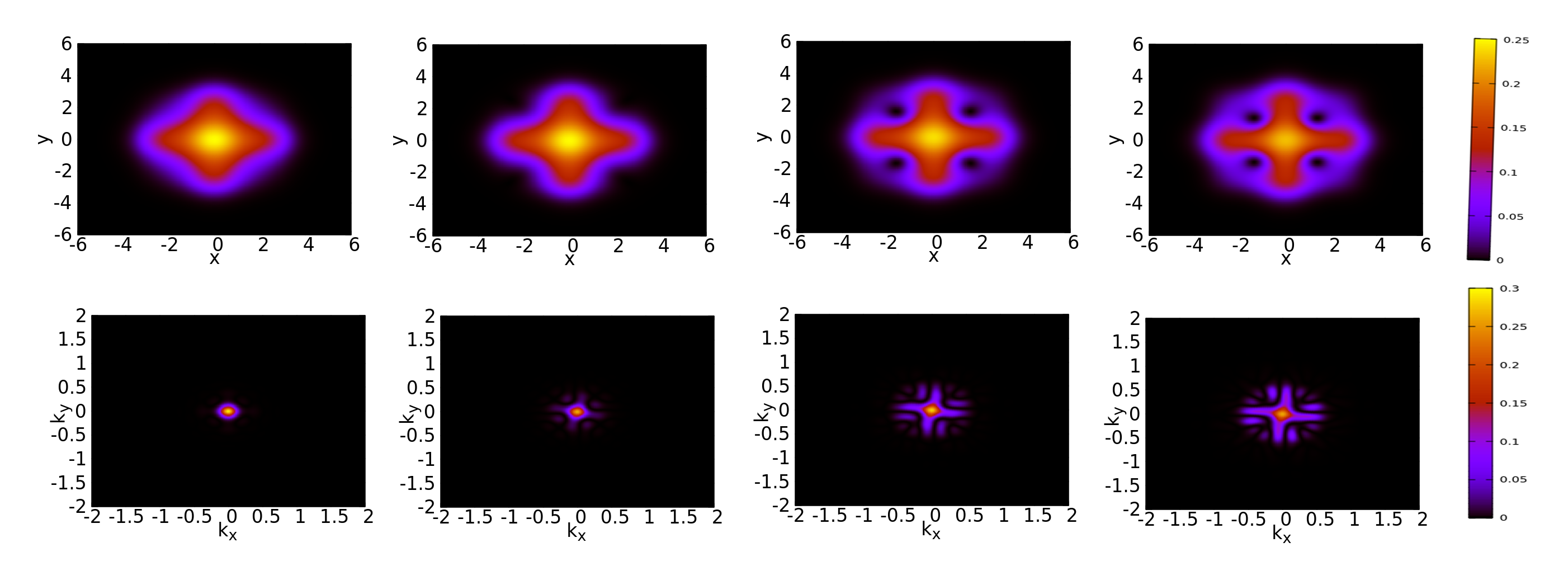}
    \caption{
    Real-space density (upper row) and corresponding momentum-space
    density (bottom row) for different angular velocities:
    $\omega=0$ (first column), $\omega=0.4$ (second column),
    $\omega=0.5$ (third column), and $\omega=0.6$ (fourth column).
    Here, $V_0=3$, $N=10000$, and $k=1$. In the presence of the
    intermediate-strength optical lattice, four vortices are generated
    progressively as the angular velocity increases. This behavior
    differs from the lattice-free case, where vortex nucleation occurs
    more abruptly. The changes in the vortex configuration are also
    reflected in the corresponding momentum-space density. The
    coordinates $x$ and $y$ are expressed in units of $l_0$, while
    $k_x$ and $k_y$ are expressed in units of $l_0^{-1}$. The units of
    $|\psi(x,y)|^2$ and $|\phi(k_x,k_y)|^2$ are $l_0^{-2}$ and $l_0^2$,
    respectively.
    }
    \label{vary_om}
\end{figure*}

\begin{figure*}
    \includegraphics[width=1.03\textwidth]{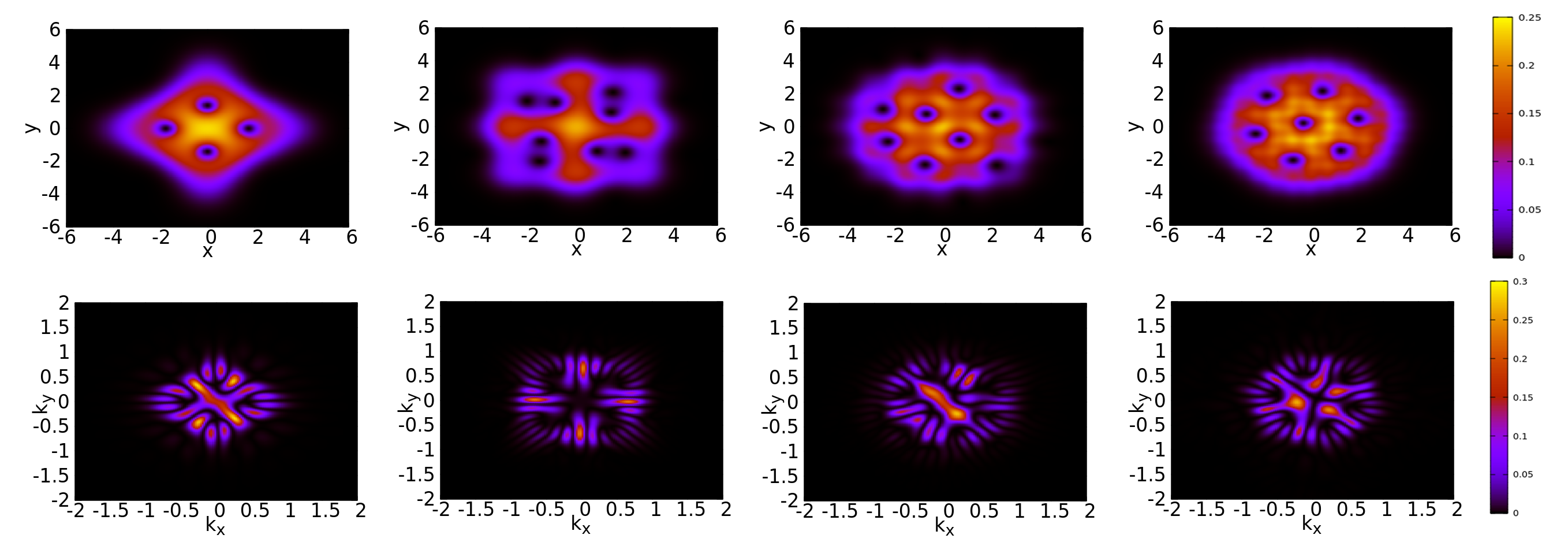}
    \caption{
    Real-space density (upper row) and corresponding momentum-space
    density (bottom row) for different optical-lattice wave vectors:
    $k=0.5$ (first column), $k=1$ (second column),
    $k=2$ (third column), and $k=4$ (fourth column).
    Here, $V_0=3$, $N=10000$, and $\omega=0.8$. For small $k$, the
    spatial variation of the optical potential is too weak to produce
    a significant modification of the condensate density. Values of
    $k$ around unity provide a spatial periodicity that strongly
    affects the condensate. For large $k$, the lattice periodicity
    becomes much smaller than the characteristic condensate length
    scale, and the optical potential approaches an effectively
    averaged contribution. Consequently, the density distribution
    approaches that of the system without a significant spatial
    modulation. The corresponding momentum-space density also changes
    with the lattice wave vector. The coordinates $x$ and $y$ are
    expressed in units of $l_0$, while $k_x$ and $k_y$ are expressed
    in units of $l_0^{-1}$. The units of $|\psi(x,y)|^2$ and
    $|\phi(k_x,k_y)|^2$ are $l_0^{-2}$ and $l_0^2$, respectively.
    }
    \label{vary_k}
\end{figure*}

\begin{figure}
    \includegraphics[width=0.48\textwidth]{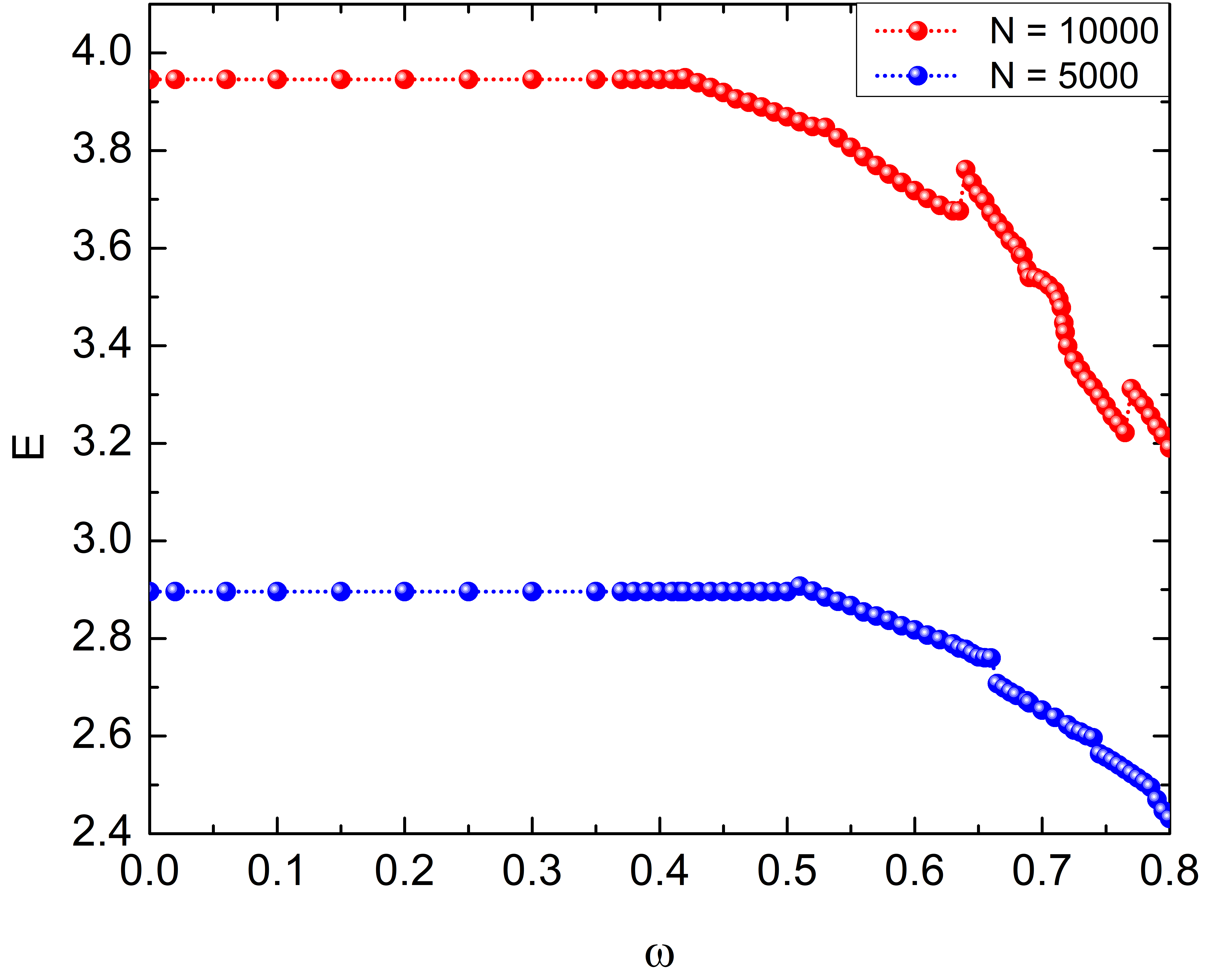}
    \caption{
    Energy per particle as a function of angular velocity in the
    absence of the optical lattice for different particle numbers.
    The energy is approximately independent of $\omega$ in the
    vortex-free regime and decreases after vortex nucleation.
    For $N=10000$, the first vortex appears at
    $\omega\simeq0.42$, whereas for $N=5000$ the corresponding critical
    angular velocity is $\omega\simeq0.51$. Thus, the critical angular
    velocity for vortex nucleation increases as the particle number
    decreases. For $N=5000$, the first, second, third, and fourth vortex
    states occur approximately in the ranges
    $0.51<\omega<0.66$, $0.66<\omega<0.74$,
    $0.74<\omega<0.79$, and $\omega>0.79$, respectively.
    For $N=10000$, the corresponding ranges for one through seven
    vortices are approximately
    $0.42<\omega<0.52$, $0.52<\omega<0.64$,
    $0.64<\omega<0.69$, $0.69<\omega<0.74$,
    $0.74<\omega<0.77$, $0.77<\omega<0.78$, and
    $\omega>0.78$, respectively. Energy gaps are observed when a new
    vortex state is generated, with the gap depending on the vortex
    configuration.
    }
    \label{evary_N}
\end{figure}

\begin{figure}
    \includegraphics[width=0.48\textwidth]{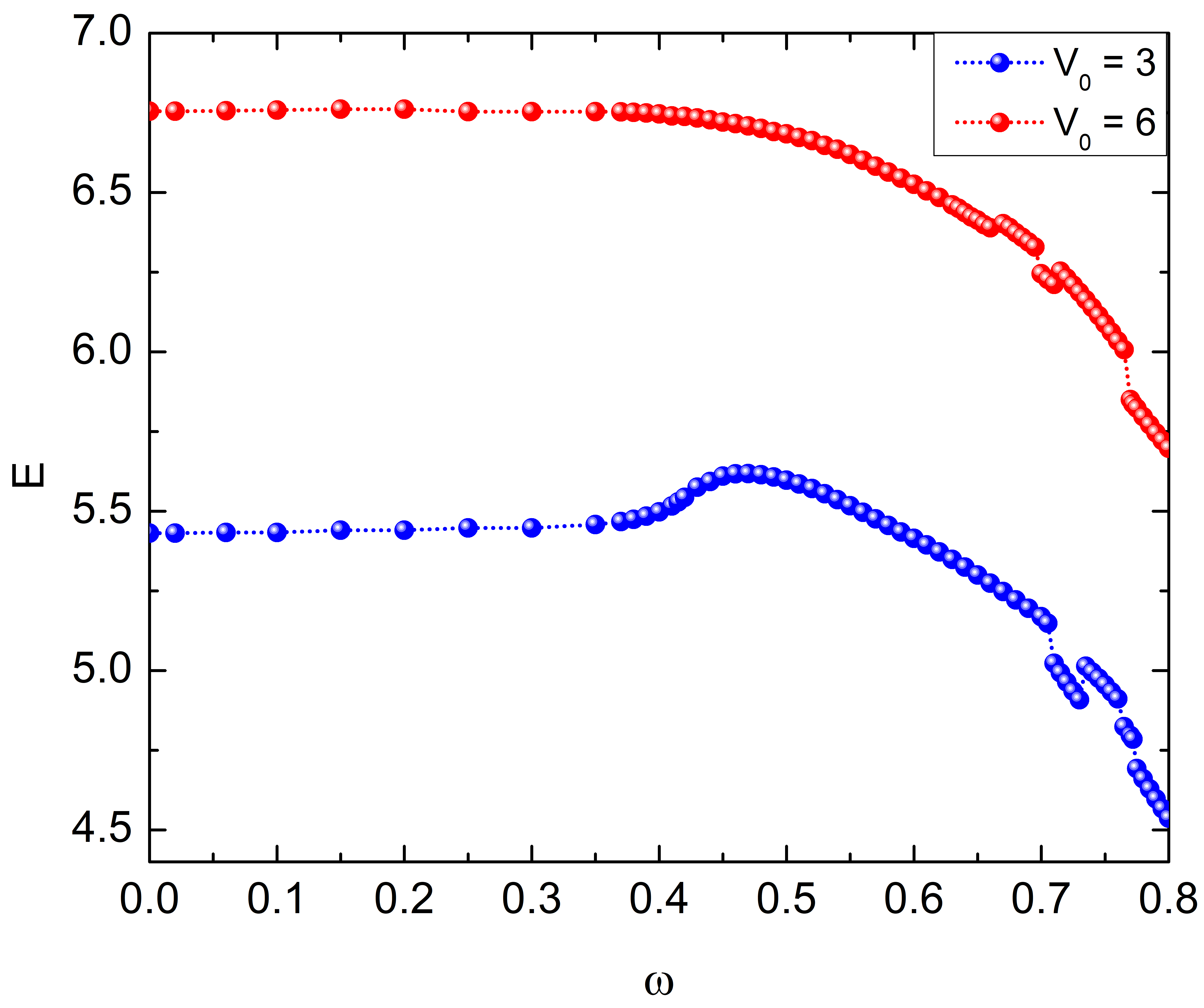}
    \caption{
    Energy per particle as a function of angular velocity in the
    presence of the optical lattice for $V_0=3$ and $V_0=6$, with
    $N=10000$ and $k=1$. The energy decreases with increasing angular
    velocity. For $V_0=3$, four vortices are generated progressively
    over the range $0.42<\omega<0.71$, followed by the formation of
    the fifth, sixth, seventh, and eighth vortices in the ranges
    $0.71<\omega<0.73$, $0.73<\omega<0.76$,
    $0.76<\omega<0.78$, and $\omega>0.78$, respectively.
    For $V_0=6$, the first four vortices are generated simultaneously.
    With further increase in $\omega$, additional vortices merge with
    the existing vortices, producing four larger vortex cores. The
    fifth, sixth, seventh, and eighth vortices merge at approximately
    $\omega>0.67$, $\omega>0.70$, $\omega>0.72$, and
    $\omega>0.77$, respectively. Energy gaps are associated with
    changes in the vortex configuration.
    }
    \label{evary_V}
\end{figure}

\begin{figure}
    \includegraphics[width=0.48\textwidth]{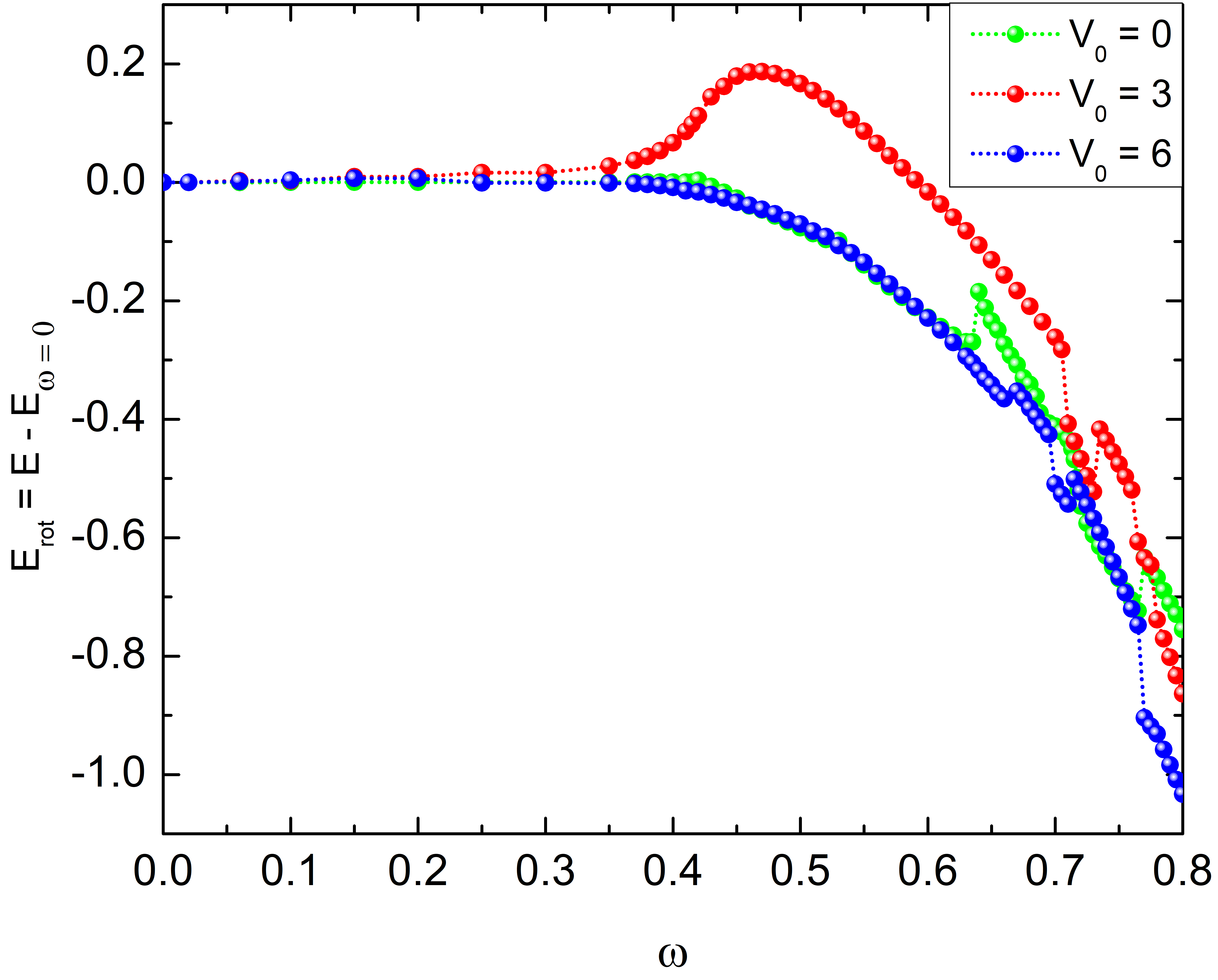}
    \caption{
    Rotational energy per particle as a function of angular velocity for different optical-lattice strengths, with $N=10000$ and $k=1$. The rotational energy, $E_{\rm rot}=E-E_{\omega=0}$, decreases with increasing $\omega$, with a progressively faster decrease for larger $V_0$. For $V_0=3$, a pronounced feature appears near $\omega\simeq0.46$ due to changes in the vortex configuration and its interaction with the optical potential. At high angular velocity, the reduction becomes particularly significant for stronger lattices; notably, near $\omega\simeq0.8$, $E_{\rm rot}$ for $V_0=6$ is substantially lower than for the weaker lattice strengths. This demonstrates the strong suppression of rotational energy by the optical lattice at high angular velocity.
    }
    \label{rot}
\end{figure}

\begin{figure}
    \includegraphics[width=0.48\textwidth]{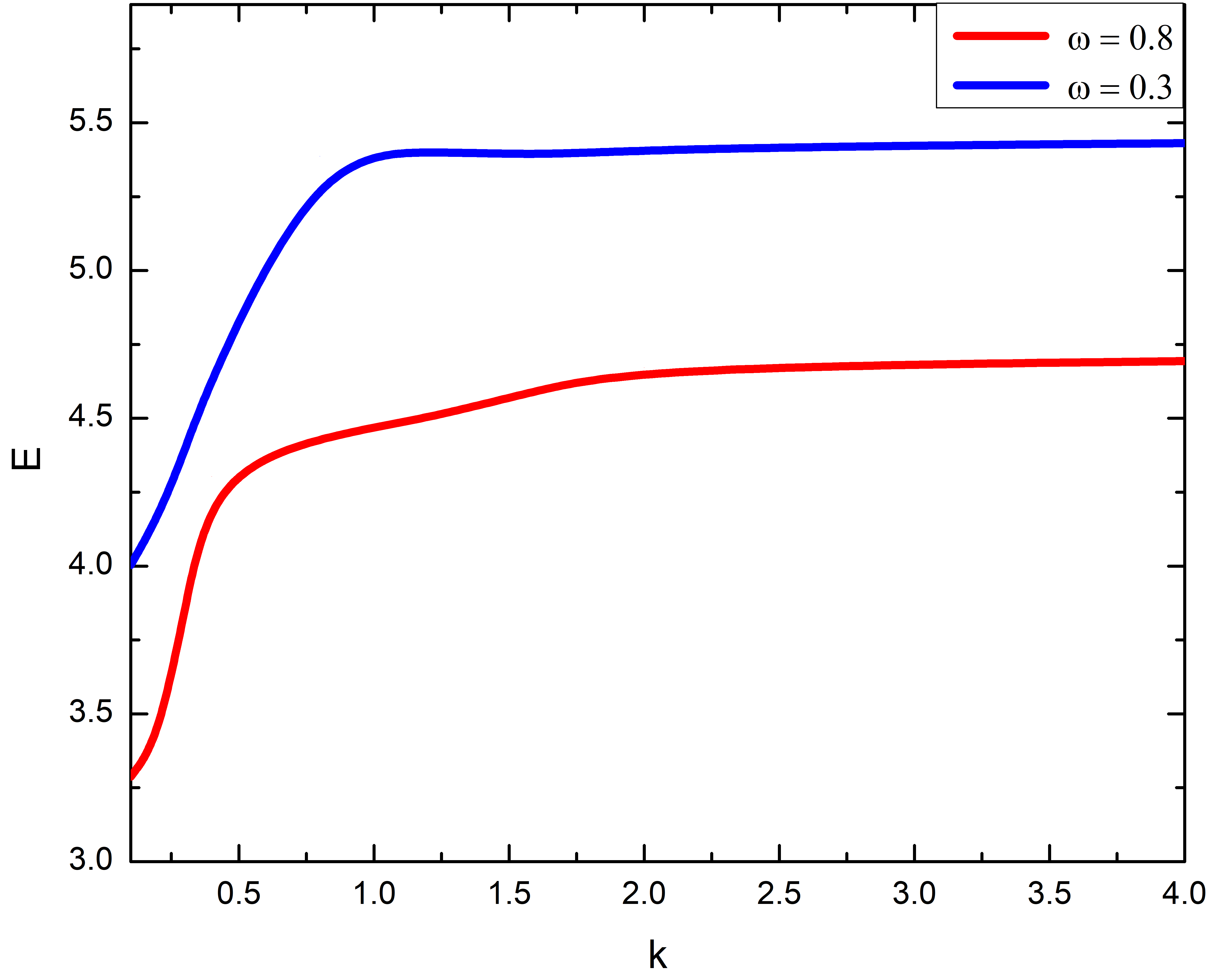}
    \caption{
    Energy per particle as a function of the optical-lattice wave vector $k$ for different angular velocities, with $N=10000$ and $V_0=3$. The energy per particle increases with increasing $k$ and gradually saturates at large $k$. The energy is lower for higher angular velocity due to the presence of vortices. At large $k$, the lattice periodicity becomes much smaller than the characteristic length scale of the condensate, and the rapidly varying optical potential is effectively averaged over the condensate, resulting in the observed saturation of the energy.
    }
    \label{evsk}
\end{figure}

\begin{equation}
    i\frac{\partial\psi}{\partial t}
    =
    \left[
    -\frac{1}{2}\nabla^2
    +\frac{1}{2}(x^2+y^2)
    +V_{\rm op}(x,y)
    +g|\psi|^2
    -\omega L_z
    \right]\psi .
    \label{GP}
\end{equation}

Here, $\psi(x,y,t)$ is the condensate wave function. The first term on the right-hand side represents the kinetic-energy contribution, while the second term corresponds to the isotropic harmonic trapping potential. The third term is the periodic optical-lattice potential given by Eq.~(\ref{Vlat}). The nonlinear term $g|\psi|^2$ describes the mean-field interaction between the condensed atoms, where $g$ is the dimensionless interaction strength. The final term, $-\omega L_z$, accounts for the rotation of the system about the $z$ axis.

The dimensionless angular-momentum operator is defined as

\begin{equation}
    L_z=-i
    \left(
    x\frac{\partial}{\partial y}
    -
    y\frac{\partial}{\partial x}
    \right).
    \label{Lz}
\end{equation}

All quantities in Eq.~(\ref{GP}) are expressed in dimensionless units. Length is measured in units of the characteristic length $l_0$ \cite{length}, energy in units of $\hbar^2/(m l_0^2)$, and time in units of $m l_0^2/\hbar$, where $m$ is the mass of the condensed atoms. The dimensionless interaction parameter $g$ is expressed in the corresponding units. The condensate wave function is normalized according to

\begin{equation}
    \int |\psi(x,y)|^2\,dx\,dy=1.
    \label{normalization}
\end{equation}

The stationary condensate configurations are obtained by solving Eq.~(\ref{GP}) in imaginary time using the split-step Crank--Nicolson method \cite{GPE1,GPE2,GPE3,avra2,GPE4,GPE5,avra1,avraj,GPE7,avra5,avra6}. For a specified set of parameters $(N,V_0,k,\omega)$, the imaginary-time evolution is continued until the condensate reaches a stationary state. The resulting real-space condensate density is given by

\begin{equation}
    n(x,y)=|\psi(x,y)|^2.
    \label{density}
\end{equation}

We use this density distribution to examine the spatial structure of the condensate and to identify the appearance and rearrangement of vortex cores. In particular, the number and spatial distribution of density depletions are analyzed as the particle number, lattice parameters, and angular velocity are varied.

\subsection*{Momentum-space density}

In addition to the real-space density, we investigate the momentum-space distribution of the condensate. This provides complementary information about how the spatial structure of the condensate is reflected in its momentum components. The momentum-space wave function is obtained from the two-dimensional Fourier transform,

\begin{equation}
   \phi(k_x,k_y)
   =
   \int_{-\infty}^{\infty}
   \int_{-\infty}^{\infty}
   \psi(x,y)
   e^{-2\pi i(k_xx+k_yy)}
   \,dx\,dy ,
   \label{fourier}
\end{equation}

and the corresponding momentum-space density is

\begin{equation}
    n_k(k_x,k_y)=|\phi(k_x,k_y)|^2.
    \label{momentum_density}
\end{equation}

The momentum-space distribution is particularly relevant in the present system because the periodic optical lattice introduces characteristic spatial Fourier components. Using

\begin{equation}
    \sin^2(kx)
    =
    \frac{1}{2}
    \left[1-\cos(2kx)\right],
\end{equation}

and an analogous expression for the $y$ direction, the lattice potential contains spatial Fourier components with characteristic wave-vector scale $2k$. Consequently, varying the lattice depth $V_0$ or the lattice wave vector $k$ can modify the distribution of momentum components of the condensate. At the same time, rotation and the associated vortex structure modify the spatial profile of the condensate and therefore its Fourier-space distribution.

We therefore use the momentum-space density as a complementary observable to characterize the response of the condensate to the combined effects of rotation and the periodic optical lattice. In the results section, the momentum-space distributions are compared for different lattice and rotational parameters to identify the corresponding changes in their structure.

\subsection*{Energy and rotational energy}

After obtaining the stationary condensate wave function, we calculate the energy of the system. Since Eq.~(\ref{GP}) is formulated in the rotating frame, the corresponding energy functional is the energy in the rotating frame. In the dimensionless units used here, the energy per particle is given by

\begin{equation}
\begin{split}
E_{\rm rot}
=
\int
\Bigg[
&\frac{1}{2}|\nabla\psi|^2
+
\left\{
\frac{1}{2}(x^2+y^2)
+
V_{\rm op}(x,y)
\right\}|\psi|^2
\\
&+
\frac{g}{2}|\psi|^4
-
\omega\psi^\ast L_z\psi
\Bigg]\,dx\,dy ,
\end{split}
\label{energy}
\end{equation}

where $\psi^\ast$ denotes the complex conjugate of the condensate wave function. The different terms in Eq.~(\ref{energy}) represent the kinetic energy, harmonic trapping energy, optical-lattice energy, mean-field interaction energy, and rotational contribution, respectively.

It is useful to separate the rotational contribution from the remaining terms. We define the non-rotating-frame contribution as

\begin{equation}
\begin{aligned}
E_{\omega = 0} = \int \bigg[ &\frac{1}{2}|\nabla\psi|^2 + \left\{ \frac{1}{2}(x^2+y^2) + V_{\rm op}(x,y) \right\}|\psi|^2 \\
&+ \frac{g}{2}|\psi|^4 \bigg] dx\,dy
\end{aligned}
\end{equation}

and the rotational contribution as

\begin{equation}
E_{\rm rot}
=
-\omega
\int
\psi^\ast L_z\psi\,dx\,dy
=
-\omega\langle L_z\rangle .
\label{Erot}
\end{equation}

Thus,

 \begin{equation}
 E=E_{\omega = 0}+E_{\rm rot}.
\label{Edecomp}
\end{equation}

This decomposition allows us to examine separately the contribution associated with rotation and the changes in the remaining energy components produced by the optical lattice. We use these quantities to investigate how the lattice depth, lattice periodicity, particle number, and angular velocity influence the energetics of the rotating condensate.

In the following sections, we first examine the spatial density and momentum-space density for different system parameters. We then analyze the vortex configurations and their rearrangement with increasing lattice strength and rotation frequency, followed by a comparison of the corresponding energy and rotational-energy contributions.

\section{Results and discussion}

We investigate the vortex configurations and energetic properties of a rotating two-dimensional Bose--Einstein condensate (BEC) in the presence of a repulsive square optical lattice. The stationary condensate wave functions are obtained by solving the rotating Gross--Pitaevskii equation [Eq.~(\ref{GP})] using the imaginary-time split-step Crank--Nicolson method. The corresponding momentum-space distributions are calculated from the two-dimensional Fourier transform given by Eq.~(\ref{fourier}). We systematically examine the dependence of the condensate density, vortex configuration, momentum-space density, and energy on the particle number $N$, lattice depth $V_0$, angular velocity $\omega$, and lattice wave vector $k$.

Figure~\ref{op} shows the two-dimensional square optical lattice used in the present calculations. Since both the lattice depth and the lattice periodicity can be varied independently through $V_0$ and $k$, respectively, the optical lattice provides a controllable spatial modulation of the condensate density. This modulation competes with the harmonic confinement, interatomic interaction, and rotational effects and consequently modifies the vortex configurations.

Figure~\ref{vary_N} shows the real-space and momentum-space densities for different particle numbers, $N=500$, $2000$, $5000$, and $10000$, in the absence of the optical lattice at fixed angular velocity $\omega=0.8$. The number of vortices increases with increasing particle number. This behavior indicates that the critical conditions for vortex nucleation depend on the density and interaction energy of the condensate. In addition to changing the number of vortices, an increase in $N$ modifies their spatial arrangement. These changes in the real-space vortex configuration are accompanied by corresponding changes in the momentum-space density. In particular, the momentum-space distribution is sensitive not only to the number of vortices but also to their spatial arrangement, since the latter modifies the Fourier components of the condensate wave function. Thus, the real-space and momentum-space distributions provide complementary descriptions of the vortex state.

The effect of the optical-lattice depth is illustrated in Fig.~\ref{vary_v} for $N=10000$, $\omega=0.8$, and $k=1$. For $V_0=1$, seven vortices are observed, whereas eight vortices are obtained for $V_0=3$. At the intermediate lattice strength $V_0=3$, the vortices exhibit a more ordered spatial arrangement due to the periodic modulation imposed by the optical lattice. Upon further increasing the lattice strength, the individual vortices begin to merge, resulting in four larger vortex cores. For sufficiently large $V_0$, the vortex structure is progressively suppressed as the density distribution becomes increasingly localized by the strong periodic potential. These observations demonstrate that the optical lattice does not merely change the number of vortices; it also provides a mechanism for modifying their spatial organization and stability. The corresponding momentum-space densities, shown in the bottom row of Fig.~\ref{vary_v}, change simultaneously with the real-space vortex configuration. The modification of the momentum distribution therefore provides a direct signature of the lattice-induced changes in the condensate structure.

The dependence on angular velocity is shown in Fig.~\ref{vary_om} for $V_0=3$, $N=10000$, and $k=1$. In the presence of the optical lattice, vortex nucleation occurs progressively as the angular velocity is increased. In the parameter range considered here, four vortices are generated gradually with increasing $\omega$. This behavior differs qualitatively from the lattice-free system, where the first vortex appears more abruptly after the critical angular velocity is reached. The optical lattice therefore modifies the vortex-nucleation process by introducing an additional spatial energy scale and preferred locations for the vortex cores. The changes in the real-space vortex configuration are also reflected in the momentum-space density. Consequently, the momentum-space distribution provides additional information about the evolution of the vortex state with angular velocity.

The dependence of the condensate density on the lattice wave vector is presented in Fig.~\ref{vary_k} for $V_0=3$, $N=10000$, and $\omega=0.8$. The lattice spacing is given by $d=\pi/k$, and hence variation of $k$ changes the spatial scale of the periodic modulation. For small $k$, the lattice varies slowly over the characteristic spatial extent of the condensate and its effect on the density distribution is comparatively weak. Around $k\simeq1$, the lattice periodicity becomes comparable to the characteristic length scale of the condensate, producing a pronounced modulation of the density and vortex configuration. For sufficiently large $k$, the lattice period becomes much smaller than the characteristic condensate length scale. The rapidly varying potential is then effectively averaged over the condensate, and the density distribution approaches that of a system in which the detailed spatial modulation of the lattice is not resolved. The results therefore indicate that the strongest structural modification occurs when the lattice periodicity is comparable to the relevant length scale of the condensate.

We next examine the energetic response of the condensate. Figure~\ref{evary_N} presents the energy per particle as a function of angular velocity in the absence of the optical lattice for $N=5000$ and $N=10000$. In the vortex-free regime, the energy is approximately independent of $\omega$, consistent with the absence of angular-momentum-carrying vortex states. For $N=10000$, the first vortex appears at a critical angular velocity of approximately $\omega_c\simeq0.42$, whereas for $N=5000$ the corresponding value is $\omega_c\simeq0.51$. Thus, within the parameter range studied, reducing the particle number increases the critical angular velocity required for the first vortex nucleation.

For $N=5000$, the one-, two-, three-, and four-vortex configurations occur approximately in the intervals $0.51<\omega<0.66$, $0.66<\omega<0.74$, $0.74<\omega<0.79$, and $\omega>0.79$, respectively. For $N=10000$, the corresponding intervals for one through seven vortices are approximately $0.42<\omega<0.52$, $0.52<\omega<0.64$, $0.64<\omega<0.69$, $0.69<\omega<0.74$, $0.74<\omega<0.77$, $0.77<\omega<0.78$, and $\omega>0.78$. The energy decreases as the system enters states with larger vortex number. The discontinuous changes or energy gaps associated with the transitions between vortex configurations reflect the change in the stationary state when an additional vortex is nucleated.

The corresponding energetic behavior in the presence of the optical lattice is shown in Fig.~\ref{evary_V} for $V_0=3$ and $6$, with $N=10000$ and $k=1$. In contrast to the lattice-free case, the energy of the vortex-free state is not strictly independent of angular velocity because the optical lattice modifies the condensate density even before vortex nucleation. For $V_0=3$, four vortices are generated progressively in the interval $0.42<\omega<0.71$. The fifth, sixth, seventh, and eighth vortices subsequently appear approximately in the intervals $0.71<\omega<0.73$, $0.73<\omega<0.76$, $0.76<\omega<0.78$, and $\omega>0.78$, respectively. Thus, the intermediate-strength optical lattice modifies the sequence of vortex nucleation compared with the lattice-free system.

For $V_0=6$, the vortex evolution is qualitatively different. The first four vortices appear simultaneously within the parameter range studied. With further increase in angular velocity, additional vortices do not remain as distinct cores but merge with the existing vortices. The fifth, sixth, seventh, and eighth vortices merge with the pre-existing vortex structures at approximately $\omega>0.67$, $\omega>0.70$, $\omega>0.72$, and $\omega>0.77$, respectively. Consequently, only four large vortex cores remain identifiable at higher angular velocities. This behavior demonstrates that increasing the lattice depth can change the mechanism of vortex evolution from sequential nucleation to vortex merging and rearrangement.

Figure~\ref{rot} shows that the rotational energy per particle decreases with increasing angular velocity, and the decrease becomes progressively steeper as the optical-lattice strength $V_0$ increases. For $V_0=3$, a pronounced peak-like feature occurs near $\omega\simeq0.46$, associated with changes in the vortex configuration and its interaction with the optical potential. At higher angular velocities, the effect of the lattice becomes increasingly significant. In particular, around $\omega\simeq0.8$, the rotational energy for $V_0=6$ is substantially lower than that for the weaker lattice strengths. This pronounced reduction is associated with the lattice-induced modification and merging of vortex structures at higher $V_0$, demonstrating that a stronger optical lattice can significantly suppress the rotational-energy contribution of the condensate at high angular velocity.

Finally, Fig.~\ref{evsk} shows the dependence of the energy per particle on the lattice wave vector $k$ for different angular velocities, with $V_0=3$ and $N=10000$. The energy per particle generally increases with increasing $k$ and approaches a saturated value at sufficiently large wave vector. This behavior is consistent with the decrease of the lattice spacing, $d=\pi/k$, as $k$ increases. When the lattice period becomes much smaller than the characteristic length scale of the condensate, the condensate effectively samples the spatially averaged contribution of the rapidly varying optical potential. The resulting saturation of the energy therefore indicates that the detailed spatial structure of the lattice becomes progressively less important at large $k$. Among the angular velocities considered, the energy is lower for $\omega=0.8$, where multiple vortices are present, reflecting the energetic contribution associated with the rotating vortex state.

\section{Conclusion}

We have investigated the vortex structure and energetic properties of a rotating BEC in the presence of an optical lattice. The results show that the momentum-space density is strongly influenced by both the number of vortices and their spatial arrangement in real space, providing a clear signature of the vortex configuration. The optical lattice promotes a more ordered vortex arrangement and modifies vortex nucleation and merging. In particular, the rotational energy decreases more strongly with increasing optical-lattice strength, with a pronounced reduction for higher $V_0$ at large angular velocity. Thus, the optical lattice provides an effective means of controlling both the vortex structure and rotational response of the BEC.

\end{document}